\documentstyle[twoside,fleqn,espcrc2,epsfig]{article}
 
\newcommand{\be}{\begin{equation}}
\newcommand{\ee}{\end{equation}}
\newcommand{\bea}{\begin{eqnarray}}
\newcommand{\eea}{\end{eqnarray}}

\newcommand{\nn}{\nonumber}
\newcommand{\msb}{\overline{\mbox{\scriptsize MS}}}

\title{Non-perturbative Renormalization of the Complete Basis of
Four-fermion Operators and B-parameters}
\author{L.~Conti\address{I.N.F.N. and Dip. di Fisica,
Universit\'a di Roma II, Rome I-00133, Italy.},
A.~Donini\address{I.N.F.N. and Dip. di Fisica, 
Universit\'a di Roma I, Rome I-00185, Italy.},
V.~Gimenez\address{Dep. De Fisica Teorica and IFIC, Univ. de Valencia,
Burjassot, Valencia E-46100, Spain.},
G.~Martinelli$^{\rm b}$,
M.~Talevi\address{Dept. of Physics and Astronomy, University of
Edinburgh, Edinburgh EH9 3JZ, UK.} and
A.~Vladikas$^{\rm a}$
\thanks{Talk presented by A.~Vladikas.}
}

\begin{document}
\begin{abstract} 
We present results on the B-parameters $B_K$, $B^{3/2}_7$ and $B^{3/2}_8$,
at $\beta=6.0$, with the tree-level Clover action. The renormalization of
the complete basis of dimension-six four-fermion operators has been performed
non-perturbatively. Our results for $B_K$ and $B^{3/2}_7$ are in reasonable
agreement with those obtained with the (unimproved) Wilson action. This is
not the case for $B^{3/2}_8$. We also discuss some subtleties arising from a
recently proposed modified definition of the B-parameters.
\end{abstract}
 
\maketitle

\section{Operator Renormalization}

In the present talk, we are interested in the determination of three
B-parameters, namely $B_K$, $B^{3/2}_7$ and $B^{3/2}_8$. $B_K$ measures the
deviation of the $\Delta S = 2$ matrix element $\langle K^0 \vert O_{\Delta S
= 2} \vert \bar K^0 \rangle$ from its value in the Vacuum Saturation
Approximation (VSA). $B^{3/2}_7$ and $B^{3/2}_8$ measure the deviation of the
$\Delta I = 3/2$ matrix elements $\langle \pi \vert O^{3/2}_{7,8} \vert \bar K
\rangle$ from their VSA values. We recall that $B_K$ is an essential
ingredient to the determination of the CP-violation parameter $\epsilon$,
whereas $B^{3/2}_7$ and $B^{3/2}_8$ are needed in the determination of the ratio
$\epsilon^\prime/\epsilon$. All three matrix elements can be computed on the
lattice from three-point correlation functions, involving the so-called
``eight" diagrams. Their renormalization has no power subtraction (involving
``eye" diagrams).

The main novelty of the present work, which is an extension of \cite{ape},
is the implementation of the Non-Perturbative Method (NPM) for the
renormalization of the corresponding operators. We have determined the operator
mixing for the complete basis of four-fermion operators with the aid of the
discrete symmetries (parity, charge conjugation and switching of flavours).
For the parity-conserving operators, relevant to this work, we have used the
following complete basis of five operators:
\bea
Q_{1,2} &=& V \times V \pm A \times A , \nonumber \\
\label{base}
Q_{3,4} &=& S \times S \mp P \times P , \\
Q_5 &=& T \times T .              \nonumber
\eea
In these expressions, $\Gamma \times \Gamma$ (with $\Gamma = V,A,S,P,T$ 
a generic Dirac matrix) stands for $\frac{1}{2}(
\bar \psi_1 \Gamma \psi_2 \bar \psi_3 \Gamma \psi_4 +
\bar \psi_1 \Gamma \psi_4 \bar \psi_3 \Gamma \psi_2) $, where 
$\psi_i,~i=1,\dots,4$ are fermion fields with flavours chosen
so as to reproduce the desired operators (see \cite{apeI,apeII}
for details): the parity-conserving component of the four-fermion 
operator $O_{\Delta S = 2}$ corresponds to $Q_1$ in our basis, whereas
the parity-conserving parts of $O_7^{3/2}$ and $O_8^{3/2}$ are
(up to numerical factors) $Q_2$ and $Q_3$. On the lattice, these operators
mix under renormalization in the following pattern
\bea
\label{eq:bk_sub}
\hat Q_1 &=& Z_{11} Q_1^{s}, \nonumber \\
\hat Q_2 &=& Z_{22} Q_2^s + Z_{23} Q_3^s , \\
\hat Q_3 &=& Z_{32} Q_2^s + Z_{33} Q_3^s . \nonumber
\eea
$Z_{11}$ and $Z_{ij}$ (with $i,j = 2, 3$) are logarithmically divergent
renormalization constants which depend on the coupling and $a\mu$. These are
renormalizations which occur also in the continuum. The subtractions
\bea
Q_1^s &=& Q_1 + \sum^5_{i=2} Z_{1i} Q_i \;\; , \\
Q_i^s &=& Q_i + \sum_{j=1,4,5} Z_{ij} Q_j \;\;\; (i=2,3) \;\; , \nn
\eea
occur on the lattice because of the chiral symmetry breaking Wilson term in
the action. The mixing coefficients in the last expressions are finite and
only depend on the lattice coupling $g_0^2(a)$. The results for all the
renormalization constants $Z_{ij}$ (computed with the NPM at several
renormalization scales $\mu$ at $\beta =  6.0$) can be found in \cite{apeI}.

The finite mixing coefficients have also been determined in \cite{jlqcd},
using Ward Identities (WI). The NPM and WI determinations are equivalent
at large enough scale $\mu$; see \cite{apeI,testa} for explicit
demonstrations. It is not true, as claimed in \cite{jlqcd} that the WI method
is theoretically more sound, off the chiral limit.
On the other hand, we have checked that the
choice of operator basis made in \cite{jlqcd} appears to give stabler results
in practice. This is not, however, a question of principle.

\begin{table*}[hbt]
\setlength{\tabcolsep}{1.5pc}
\newlength{\digitwidth} \settowidth{\digitwidth}{\rm 0}
\catcode`?=\active \def?{\kern\digitwidth}
\caption{
B-parameters for $\Delta S=2$ and $\Delta I =3/2$ operators at the
renormalization scale $\mu = a^{-1} \approx 2$ GeV. All results are in
the $\msb$ renormalization scheme (with the dimensional regualrization
shown in the third column).  $\hat B_K$ is the RGI B-parameter
(obtained by multiplying $B_K$ by its Wilson coefficient).
}
\label{tab:bpar}
\begin{tabular*}{\textwidth}{@{}l@{\extracolsep{\fill}}lllll}
\hline
            & NPM                & NDR & 0.66(11) & this work       \\
$B_K$       & BPT                & NDR & 0.65(11) & this work       \\
            & BPT $q^\ast = 1/a$ & NDR & 0.74(4)  & \cite{gbs} \\
\hline
            & NPM                & NDR & 0.93(16) & this work \\
$\hat B_K$  & BPT                & NDR & 0.92(16) & this work \\
\hline
            & NPM                  & NDR &  0.72(5)  & this work       \\
$B_7^{3/2}$ & BPT                  & DRED & 0.65(2)  & this work       \\
            & BPT $q^\ast = 1/a$   & NDR & 0.58(2)  & \cite{gbs} \\
            & BPT $q^\ast = \pi/a$ & NDR & 0.65(2)  & \cite{gbs} \\
\hline
            & NPM                  & NDR & 1.03(3)  & this work       \\
$B_8^{3/2}$ & BPT                  & DRED & 0.71(2)  & this work       \\
            & BPT $q^\ast = 1/a$   & NDR & 0.81(3)  & \cite{gbs} \\
            & BPT $q^\ast = \pi/a$ & NDR &0.84(3)  & \cite{gbs} \\
\hline
\end{tabular*}
\end{table*}

\section{The Definition of $B_K$}

The standard definition of $B_K$ is given by
\bea
B_K(\mu) &=& \frac{\langle \bar K^0 \vert \hat O_{\Delta S = 2} (\mu) 
\vert K^0 \rangle}{\langle \bar K^0 \vert O_{\Delta S = 2} \vert K^0 \rangle
_{VSA}} \nonumber \\
&=& \frac{\langle \bar K^0 \vert \hat O_{\Delta S = 2} (\mu) 
\vert K^0 \rangle}{\frac{8}{3} f_K^2 m_K^2 }
\eea
Note that the operator $\hat O_{\Delta S = 2}$ in the numerator is
renormalized. Thus the numerator of the above ratio is a $\mu$-dependent
quantity, whereas the denominator is a physical one.
Thus defined, $B_K$ scales with $\mu$ in the same way as
$\hat O_{\Delta S = 2}$. In \cite{jlqcd}, the following modified definition has
also been used:
\bea
B_K^{\prime}(\mu) &=& \frac{\langle \bar K^0 \vert \hat O_{\Delta S = 2} (\mu) 
\vert K^0 \rangle}{\langle \bar K^0 \vert \hat O_{\Delta S = 2} (\mu)
\vert K^0 \rangle_{VSA}} \nonumber \\
&=& \frac{\langle \bar K^0 \vert O_{\Delta S = 2}^s (a) 
\vert K^0 \rangle}{\langle \bar K^0 \vert O_{\Delta S = 2}^s (a)
\vert K^0 \rangle_{VSA}}
\eea
In other words, each operator subtraction in the renormalization of
$ \hat O_{\Delta S = 2} (\mu) $ is vacuum-saturated in the denominator.
This modified definition results in a statistically stabler signal.
In \cite{jlqcd}, both $B_K$ and $B_K^{\prime}$ were measured
at several $\beta$ values and the results, after being extrapolated at
zero lattice spacing, were found to be compatible. However, it can be shown
\cite{ms} that this definition has serious shortcomings. The problem
lies with the denominator, which, up to terms proportional to the lattice
spacing, behaves like
\be
\frac{1}{Z_A^2} \frac{8}{3} f_K^2 m_K^2 +
Z_P (g_0^2) \vert \langle 0 \vert P(a) \vert K^0 \rangle \vert ^2
\ee
where $Z_A$ and $Z_P (g_0^2)$ are the renormalization constants of the axial
current $A_\mu$ and the pseudoscalar density $P$.
The numerator has the correct chiral behaviour, but that of the denominator
is spoiled by $\cal{O} (g_0^2)$ terms. These terms could be eliminated by
extrapolating to the continuum limit $a \rightarrow 0$ before taking the
chiral limit (in the continuum limit the denominator of $B_K^\prime$ reduces
to that of $B_K$). But this is not possible, as the numerator diverges in this
limit. A possible remedy of this problem would be a further
modification of the definition of $B_K$:
\be
B_K(\mu)^{\prime\prime} = \frac{Z_O(a\mu)}{Z_A^2} B_K^\prime (\mu)
\ee
which has a finite numerator (also in the continuum limit) and a denominator
\be
\frac{8}{3} f_K^2 m_K^2 +
\frac{Z_P}{Z_A^2} (g_0^2) \vert \langle 0 \vert P(a) \vert K^0 \rangle \vert ^2
\ee
The last term scales like $[g_0^2]^{3/11}$, since $Z_P(g_0^2) \sim g_0^2$
and $P(a) \sim [g_0^2]^{-8/11}$. Thus it vanishes very
slowly in the continuum limit, and cannot be removed by a linear extrapolation
in $a$ (as suggested in \cite{jlqcd}).

\section{Results}

Our results have been obtained with the tree-level Clover action, at
$\beta =6.0$ in the quenched approximation. The matrix elements have been
computed on an
$18^3 \times 64$ lattice (460 configurations), whereas the non-perturbative
renormalization (based on the computation of the matrix elements of the
operators between quark states) has been performed on a $16^3 \times 32$
lattice (100 configurations). In table 1 we present our results and compare
them to those of \cite{gbs}, also obtained at $\beta = 6.0$, but with
the (unimproved) Wilson action and with the operator renormalization done
in Boosted Perturbation Theory (BPT), which involves an ``optimal"
renormalization scale $q^\ast$. We also show our preliminary analysis
in BPT, for comparison (our BPT prescription does not make use of $q^\ast$; see
\cite{apeII} for details). Any differences arising from the use of
two regularization schemes (NDR and DRED) in $\msb$
are small and are properly accounted for in \cite{apeII}.

Our results for $B_K$, obtained with the NP and the BPT
renormalization of the operators are in perfect agreement.
With a larger statistical error, our $B_K$ value also 
agrees with those of \cite{gbs}.
Also for $B^{3/2}_7$, our NPM and BPT values are in good agreement and fully
compatible with the results of \cite{gbs} (for large enough $q^\ast$).
We find, instead, a large difference between our NPM and BPT estimates of
$B_8^{3/2}$. Our value obtained with BPT is close to that of
\cite{gbs}, where the Wilson action was used. The  NPM estimate,
instead, is
in disagreement with any value obtained in BPT (either with the Wilson or the
Clover action and for several boosting variants). We believe that the
difference between our NPM estimate and that of \cite{gbs} is due
to the NPM used in the former result, rather than the implementation
of different actions (Clover and Wilson respectively).
The increase in the NPM value of $B_8^{3/2}$ is of
great phenomenological interest, since it may induce a considerable decrease
of the ratio $\epsilon^\prime/\epsilon$.


\begin{thebibliography}{10}

\bibitem{ape}
A.~Donini {\em et al.}, Phys.~Lett. {\bf B360} (1996) 83;
M.~Crisafulli {\em et al.}, Phys.~Lett. {\bf B369} (1996) 325;
A.~Donini {\em et al.}, Nucl.~Phys. {\bf B} (Proc.~Suppl.) {\bf 53} (1997) 883;
M.~Talevi, hep-lat/9705016.
\bibitem{apeI}
A.~Donini, V.~Gimenez, G.~Martinelli, M.~Talevi and A.~Vladikas,
prep. ROME1-1181/97  in preparation.
\bibitem{apeII}
L.~Conti, A.~Donini, V.~Gimenez, G.~Martinelli, M.~Talevi and A.~Vladikas,
prep. ROME1-1180/97  in preparation.
\bibitem{jlqcd}
JLQCD Collaboration, S. Aoki {\em et al.}, hep-lat/9705035;
Nucl.~Phys. {\bf B} (Proc. Suppl.) {\bf 53} (1997) 349.
\bibitem{testa}
M.~Testa, these proceedings.
\bibitem{ms}
G.~Martinelli and S.~Sharpe, private communication.
\bibitem{gbs} 
T.~Bhattacharya, R.~Gupta and S.~Sharpe, Phys.~Rev. {\bf D55} (1997) 4036.

\end{thebibliography}
\end{document}